\documentclass{article}

\usepackage{arxiv}

\usepackage{doi}
\usepackage{natbib}
\usepackage{hyperref}       % hyperlinks
\usepackage{url}            % simple URL typesetting
\usepackage{booktabs}       % professional-quality tables
\usepackage{amsfonts}       % blackboard math symbols
\usepackage{nicefrac}       % compact symbols for 1/2, etc.
\usepackage{microtype}      % microtypography
\usepackage{xcolor}         % colors
\usepackage[utf8]{inputenc} % Required for inputting international characters
\usepackage[T1]{fontenc} % Output font encoding for international characters
\usepackage{palatino} 
\usepackage{pslatex}
\usepackage[ruled,linesnumbered,vlined]{algorithm2e}
\usepackage{pdfpages}
\usepackage{times}
\usepackage{amsmath}
\usepackage{amsthm}
\usepackage{amssymb}
\usepackage{diagbox}
\usepackage{csquotes}
\usepackage[shortlabels]{enumitem}
\usepackage{float}
\usepackage{graphicx}
\usepackage{subcaption}
\usepackage[skip=6pt plus1pt, indent=23pt]{parskip}

\title{Does Calibration Affect Human Actions?}

\author{
    Meir Nizri \\
    Ariel University
    \And
    Amos Azaria \\
    Ariel University
    \And
    Chirag Gupta \\
    Carnegie Mellon University
    \And
    Noam Hazon \\
    Ariel University
} 

\begin{document}
\maketitle

\begin{abstract}
Calibration has been proposed as a way to enhance the reliability and adoption of machine learning classifiers. We study a particular aspect of this proposal: how does calibrating a classification model affect the decisions made by non-expert humans consuming the model's predictions? We perform a Human-Computer-Interaction (HCI) experiment to ascertain the effect of calibration on (i) trust in the model, and (ii) the correlation between decisions and predictions.
We also propose further corrections to the reported calibrated scores based on Kahneman and Tversky's prospect theory from behavioral economics, and study the effect of these corrections on trust and decision-making.
We find that calibration is not sufficient on its own; the prospect theory correction is crucial for increasing the correlation between human decisions and the model's predictions. While this increased correlation suggests higher trust in the model, responses to ``Do you trust the model more?" are unaffected by the method used.
\end{abstract}

% -----------------------------------------------------------

\section{Introduction}

% Importance of ML models and the danger in high-risk decisions
Machine learning (ML) models often assume the role of assistants rather than sole decision-makers, aiding humans in making the final decision that leads to a tangible action. However, in some areas where decisions are made, such as weather forecasts, medical diagnoses, risk assessments, fraud detection, and financial forecasting, it is essential that the model not only provides a predicted class but also associates its classification with a probability. In a classification setting, confidence takes the form of reporting a probability associated with the predicted class or probability. 

% Examples and explanation of why we need probabilities
As a simple motivating example, consider a couple planning an outdoor wedding. It is insufficient for a model to predict that there will be no rain on their wedding day; the model must also provide a probability. It is quite likely that even if this probability is less than 50\%, but not negligible (e.g., 30\%), the couple will organize their wedding indoors. That is, the couple is not concerned only with whether the model predicts rain or not; they are interested in the probability of rain.

% Confidence score
Indeed, classification models are typically inherently probabilistic; that is, their raw outcome is a probability (or a set of probabilities) that is translated into a class prediction. It is common practice to present a probability to the user alongside each prediction; this probability is often referred to as a \emph{confidence score}. When confidence scores are reported, the human decision-maker consuming them must use them in a way that is consistent with what the machine learning expert had in mind when designing the model.

% Calibration importance
The first consideration we can make is whether the model is calibrated. That is, does the probability of the predicted class align with the confidence score provided? Many out-of-the-box approaches lead to classifiers that produce over-confident predictions. For instance, a model might predict a rainy day with a $90\%$ probability. However, if we analyze all the model predictions calling for rain with a $90\% $ probability, only $70\%$ might correspond to actual rainy days.
Specifically, it is well known that modern neural networks suffer from over-confident predictions \cite{guo2017calibration}. 

% Well and poor calibrated models
A well-calibrated model should produce a probability that accurately reflects the true likelihood that the event will occur. For example, if a well-calibrated model predicts a rainy day with an 80\% probability, it should rain on approximately 80\% of the days with such a prediction. We note that if, in practice, either a higher or lower percentage of days are rainy than predicted, the model is not considered calibrated. 
Several papers have proposed methods to produce classifiers that are calibrated; see \citet[Chapter 1]{gupta2023post} for a recent review of the literature. For completeness, we also review some calibration methods and their quantitative evaluation metrics in Appendix \ref{subsec:cal-met} and \ref{subsec:cal-eval} respectively.

% Human evaluation of calibration
However, despite the existence of many different calibration methods, which compose an entire field in machine learning, little is known about how \emph{humans} react to a calibrated model in comparison to the original uncalibrated model. A human-focused evaluation of calibration is natural to ask for. Such an evaluation offers multiple benefits as noted next.
First, human evaluation provides a means to assess the real-world impact and effectiveness of the model. While traditional evaluation metrics offer quantitative insights, human subjects can provide qualitative feedback that captures nuances and contextual considerations that may be missed by automated assessments alone \citep{bostrom2018ethics}. 
Second, human evaluation allows for assessing the usability and user experience of the model. Feedback from humans can provide valuable insights into how well the model meets the needs and expectations of users \citep{Kaplan2019SiriSI}.
Furthermore, human evaluation helps to ensure that the model aligns with ethical and legal standards. It allows for the examination of the fairness, transparency, and accountability of the model, as well as its compliance with regulations and guidelines \citep{mittelstadt2016ethics}.

\subsection{An additional layer of Kahneman and Tversky's prospect theory}

% Calibration based on prospect theory
In this paper, we introduce an additional layer on top of the existing calibration methods, based on the \emph{prospect theory} from behavioral economics \citep{kahneman1979prospect, kahneman1992Advances}. 
%
% Explain prospect theory
Prospect theory provides a behavioral or psychological framework that seeks to explain how individuals perceive and evaluate probabilities of potential gains and losses. We posit that considering prospect theory is crucial since it is the perceived probability that influences individuals' decision-making and trust in the system, and not the actual reported probability. 
Prospect theory uses a weighting function to describe how people subjectively weigh probabilities. In a nutshell, events with very low probabilities (i.e., close to 0) are often perceived as more likely than they truly are, while events with very high probabilities (i.e., near 1) are perceived as less likely than they truly are.
%people are more motivated to avoid losses than to achieve gains. Hence, when confronted with losses people often exhibit risk-seeking behavior, whereas, in the face of gains, they tend to be risk-averse, aiming to avoid losses and maintain their current situation. 

% How our calibration method works
In our approach, we propose transforming each calibrated probability based on the inverse of the prospect theory weighting function to better match user perceptions. Thus, by using the inverse function, we can derive a probability that users perceive as matching the original prediction. For instance, if individuals perceive an event with a reported probability of $90\%$ as actually having an $80\%$ chance of occurring, then for an actual probability of $80\%$, we would report it as $90\%$ to align with their perception.

\subsection{Experimental method and summary of findings}
% Experiment general explanation
We performed a human study to evaluate our prospect theory-based calibration method on two distinct settings. The first domain is rain forecasting, and the second is loan approval. Both domains are commonly associated with probabilistic predictions and are easily interpretable by a general audience.

We used a neural network as the uncalibrated model. We tested several calibration methods and selected the \emph{isotonic regression} method, which performed better than other methods. 
We compare our method, which adds the prospect theory correction on top of the calibrated model, to four baselines. Namely, the uncalibrated model, the model calibrated with isotonic regression, the prospect theory correction directly over the uncalibrated model, and a random method using our method's probabilities. % our method with the reported outcome drawn uniformly from $\{0,1\}$. 
%The last baseline corresponds to changing the outcome instead of the reported probability, whereas the other three correspond to changing the reported probability.

In the rain‐forecasting domain, we presented each participant with a rain prediction system. Following each prediction, participants were asked to indicate, assuming they are planning an outdoor activity for that day, how likely they are to cancel the activity. Additionally, they were requested to rate their level of trust in the predictive model.
In the loan‐approval domain, each scenario began with detailed borrower and loan information, and participants were asked to indicate the probability that they would approve the loan. We then revealed the system’s prediction and allowed participants to revise their initial decision. Finally, participants provided a trust rating for the predictive model.

% Surveys results
Our results, from both domains, show that there is no apparent difference in the level of trust reported by the participants. Nonetheless, in the correlation between participants' decisions and the models' predictions, we observed a significant difference. Our method resulted in a significantly higher correlation than all other baselines. These results indicate the utility of incorporating prospect theory into calibration methods.

% This survey seeks to understand how non-experts people without specialized knowledge in fields like mathematics, computer science, statistics, or meteorology perceive machine learning. This group is becoming increasingly involved in using machine learning, but their limited technical expertise makes it difficult for them to assess the quality of predictions effectively. Previous studies have shown that laypeople often have difficulty interpreting numerical data and may continue to trust the recommendations of a machine learning model, even when those predictions are shown to be inaccurate \citep{peters2006numeracy, reyna2008numeracy}.

% Summery
To summarize, the main contribution is providing a novel approach that incorporates the principles of prospect theory into existing calibration methods. Furthermore, it describes an evaluation with human participants to validate the effectiveness of this approach and investigate the general impact of model calibration on humans.

% -----------------------------------------------------------
%
\section{Related work on decision-making and calibration}

% What affects people to trust the ML model
The relationship between calibration, AI-assisted decision-making, and trust appears to be relatively under-explored, but we review several relevant papers. \cite{Rechkemmer2022When} examine in their survey how the following performance indicators affect people's trust in the model: the model stated general accuracy (on the test group), the model's actual accuracy (after observing several of its predictions), and the level of confidence score accompanying each prediction of the model. Their findings revealed that the level of confidence had minimal impact on trust, whereas the stated accuracy and actual accuracy of the model had a more substantial influence. 
In another study conducted by \cite{Trust2019Kun}, they examined the circumstances under which people trust AI systems and how they perceive the system's predictions when they are required to collaborate with it in making decisions. Their survey results unveiled that for systems with $70\%$ accuracy or higher, people tended to increase their trust in the system after a series of experiments. Conversely, in systems with lower accuracy people's trust is decreasing and they rely more on their own judgments.

% How to calibrate human trust 
\cite{effect2020Zhang} examined through a survey whether the presentation of a confidence score and local explanations, that accompany each prediction of the model, influences the calibration of people's trust in the model. Trust calibration involves adjusting the level of trust people place in the model to better align with its actual performance. This process is crucial because, in many cases, users' expectations or intuitions about the model's performance may not accurately reflect its true capabilities. The results show that presenting the model level of confidence does help to calibrate trust, but it alone is not sufficient to improve shared decision-making. Other factors also play a role, such as whether the person brings enough unique knowledge to supplement the model's errors. In addition, it is shown that local explanations do not create a noticeable effect on trust calibration.
\cite{Investigating2022Diniz} also conducted a study aimed at calibrating people's trust in a machine-learning model. They tested whether adding class probabilities as a confidence score, when classifying animals in images, will help to calibrate people's trust. Their results revealed that incorporating class probabilities, especially in instances where distinguishing between different animals was challenging, did not lead to a significant improvement and even increased skepticism in some cases.

% Does calibration affect human trust
It is commonly believed that the confidence value should be a well-calibrated estimate of the probability that the predicted label matches the true label \citep{gneiting2007probabilistic}. However, \cite{vodrahalli2022uncalibrated} has suggested that AI models that report confidence values that do not align with the true probabilities can encourage humans to follow AI advice. They demonstrate through a survey with humans that presenting AI models as more confident than they actually are can enhance human-AI performance in terms of accuracy and confidence of the humans in their final decision. This finding is validated across different tasks and supported by simulation analysis.
In our work, we add an additional layer based on the prospect theory over a calibrated model and show an improvement in people's decision-making. Furthermore, we focus on situations in which the model must predict a probability rather than predicting a class and a confidence value. Finally, we test several baselines, including an uncalibrated model, a calibrated model, the uncalibrated model with the prospect theory, and our method with random outcomes.

% (Cumulative) Prospect Theory (including what you used for the $\gamma$ 0.71)

% application that used prospect theory

% -----------------------------------------------------------

\section{The prospect theory correction}
\label{sec:pt}
% Recall what is prospect theory
According to prospect theory, people evaluate probabilities based on a reference point, typically the status quo or their current situation. Their evaluations are influenced by whether an outcome is perceived as a gain or a loss relative to the reference point. People often exhibit risk-seeking behavior when facing gains and risk-averse behavior when facing losses, as they strive to avoid losses and maintain the reference point. This phenomenon is known as \textit{loss aversion} \citep{kahneman1979prospect, kahneman1992Advances}. 

% Why to use it in model calibration
Incorporating prospect theory into model calibration can improve the accuracy and realism of the model's predictions, making it more effective in capturing human behavior.
Since prospect theory accounts for the biases and heuristics that individuals rely on when making decisions, incorporating these biases can better simulate and predict how people respond to different scenarios. 

% Our Method explanation (weighting function, inverse)
Prospect theory uses a non-linear weighting function to describe how people subjectively weigh probabilities. The standard weighting function is
\begin{equation}
w(p) = \frac{p^\gamma}{(p^\gamma+(1-p)^\gamma)^\frac{1}{\gamma}}, \label{eq:pt-forward}
\end{equation}
with the parameter $\gamma \in (0, 1]$ describing the amount of over and under-weighting. The weighting function has different $\gamma$ values for gains and losses. 
Typically, $\gamma$ is determined empirically through experiments that assess how people subjectively perceive and weigh probabilities. The value of $\gamma$ can vary depending on factors such as the context of the task, individual differences, and cultural influences \citep{kahneman1992Advances}. To guarantee a monotone probability weighting function, the value of $\gamma$ has to be larger than $0.279$ \citep{Cumulative2006Marc, Monotonicity2008Ingersoll, DEGIORGI2012951}. This function captures the diminishing sensitivity to gains and increasing sensitivity to losses. 

In our approach, we propose to transform each calibrated probability using the inverse of the weighting function. Since there is no simple closed-form formula for the inverse of the weighting function, we compose the following approximation:
\begin{equation}
w^{-1}(p) \approx \frac{p^\frac{1}{\gamma}}{(p^\frac{1}{\gamma}+(1-p)^\frac{1}{\gamma})^\frac{1}{\gamma}}.\label{eq:pt-backward}
\end{equation}

\begin{figure}[htb]
    \begin{center}
    \includegraphics[width=0.65\linewidth]{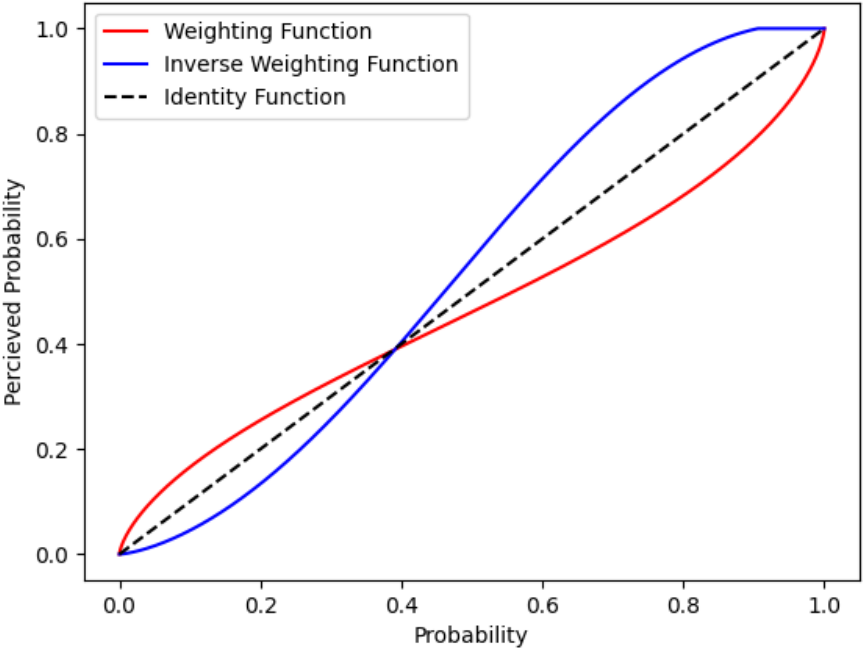}
    \end{center}
    \caption{Prospect theory uses a weighting function to transform reported probabilities into perceived probabilities. The red curve represents the weighting function \eqref{eq:pt-forward} that maps reported probabilities on the X-axis to perceived probabilities on the Y-axis; the blue curve is its approximate inverse \eqref{eq:pt-backward}. Plots are with $\gamma = 0.71$.}
    \label{fig:weightFunc}
\end{figure}

Figure \ref{fig:weightFunc} illustrates the relation between the weighting function \eqref{eq:pt-forward} and our approximate inverse function \eqref{eq:pt-backward} for $\gamma = 0.71$.

To evaluate how accurately our proposed function approximates the inverse of the weighting function, we calculated the mean absolute error (MAE) between all probabilities $P \in \{0, 1, 2, \dots, 100\}$ and $w^{-1}(w(P))$ to be $0.963$, which corresponds to a deviation of less than one percent. Additionally, the mean squared error (MSE) is $0.022$.

% -----------------------------------------------------------------

\section{Experimental design}

\subsection{Overview}

We evaluate our approach in two human-subject experiments, namely the rain forecasting domain and the loan approval domain. 
For generating predictions, we train a multi-layer perceptron (MLP) with four hidden layers and a sigmoid output. For each domain, we first compare four popular post-hoc calibration schemes—Platt scaling, temperature scaling, Bayesian binning, and isotonic regression. In both domains, isotonic regression achieved the best trade-off between calibration error and accuracy (see Appendix A and Table \ref{tbl:calibrationEval} and Table \ref{tbl:calibrationEval2}). Consequently, all subsequent human evaluations used the isotonic-regressed network as the calibrated model. Our inverse prospect-theory transformation used a $\gamma$ of $0.71$, the value empirically fitted for U.S. respondents by \citet{Estimating2017Rieger}.

Overall, we tested $5$ methods as our system's prediction:
\begin{enumerate}
    \item Uncalibrated: the MLP model described above without any calibration method.
    \item Calibrated version of the MLP model, with calibration done using isotonic regression.
    \item PT-calibrated: each calibrated probability is transformed using the prospect theory method as described in section \ref{sec:pt}. %We used $\gamma = 0.71$, which is the fitting $\gamma$ value for USA residents according to a survey conducted by \cite{Estimating2017Rieger}.
    \item PT-uncalibrated: applying the prospect theory correction to the uncalibrated model.
    \item Random: the probabilities presented to participants in this method are similar to those in the PT-calibrated method, except that the actual outcome for each prediction (whether it rains or not) is selected randomly. This method is included to ensure that any differences between the PT-calibrated method and the other models are not solely due to the reported probabilities, but are in relation to the true predictions.
\end{enumerate}

% Hence, every participant was exposed to five model variants that differed only in the post-processing of the raw network scores but were otherwise identical.

The survey workflow included a short task-specific introduction. Each participant encountered 20 independent scenarios drawn at random from the test set. For every scenario, the procedure comprised three screens. 
%First, the participant viewed the model’s probability estimate and recorded an action tendency on a five-point Likert scale. 
%Second, the true outcome was revealed. 

In the rain forecasting domain, on the first screen, we present the system's prediction for the likelihood of rain on that day. Participants are asked to assume they are planning an outdoor activity on that day and, based on the system's prediction, indicate how likely they are to cancel this activity. Participants can choose from five options on a Likert scale \citep{joshi2015likert}, ranging from ``not at all" (1) to ``very much" (5).
On the second screen, the real outcome (whether it rained or not) is displayed along with an appropriate visual image.

In the loan approval domain, on the first screen, we present borrower-loan details and ask the participants how likely they are to approve the loan on a scale from 1 (not likely) to 5 (very likely). Then, they observe the system's prediction likelihood that the borrower will repay the loan, and they have the option to update their decision. The second screen presents the real loan outcome (repaid or defaulted) accompanied by a relevant visual representation."

The final screen is identical in bot domains. On the final screen, the participants are required to recall both the prediction and the outcome. This step ensures that participants are actively engaged in considering the predictions and outcomes rather than simply advancing to the next page. Finally, they rate the model’s trustworthiness on a seven-point Likert scale, spanning from ``not at all" (1) to ``very much" (7).

We note that participants were asked to specify their degree of trust in the system's predictions and to take a domain-specific action based on those predictions. This design was chosen to estimate participants' perceived reliability of the system. We hypothesize that while people may find it challenging to articulate their exact level of trust, observing the correlation between their actions and the system's predictions might offer a more accurate indication of their trust. We used different Likert scales during the experiment (1-5 and 1-7) to ensure that the participants think and answer each question separately (so that the answer to one question will have a lesser effect on the other question).

Demographic information was collected at the end. All studies were run on Amazon Mechanical Turk with U.S. workers whose approval rate was at least 99\%; no master's qualification was required. Participants received a base payment and could earn bonus incentives that differed slightly between domains (details below).

\begin{table}[htb]
\begin{center}
    \begin{tabular}{ccccccc}
        \toprule
            Model & Acc$\uparrow$ & F1$\uparrow$ & ECE$\downarrow$ & NLL$\downarrow$ & Brier$\downarrow$ \\
        \midrule
            Uncalibrated model & 0.794679 & 0.787876 & 0.088844 & 0.5448 & 0.153413 \\
            Platt scaling & 0.796196 & 0.794154 & 0.038459 & 0.459308 & 0.146689 \\
            Isotonic regression & \textbf{0.79724} & 0.796942 & \textbf{0.007745} & \textbf{0.447345} & \textbf{0.142313} \\
            Binning with PS & 0.793212  & \textbf{0.801376} & 0.008592  & 0.449277 & 0.142469 \\
        \bottomrule \\
        \end{tabular}
\end{center}
\caption{Calibration methods comparison on the rain-forecasting dataset.}
\label{tbl:calibrationEval}
\end{table} 

\begin{table}[htb]
\begin{center}
    \begin{tabular}{ccccccc}
        \toprule
            Model & Acc$\uparrow$ & F1$\uparrow$ & ECE$\downarrow$ & NLL$\downarrow$ & Brier$\downarrow$ \\
        \midrule
            Uncalibrated model & 0.805468 & \textbf{0.818877} & 0.063677 & 0.442009 & 0.142277 \\
            Platt scaling & 0.804094 & 0.811863 & 0.024666 & 0.425698 & 0.137868 \\
            Isotonic regression & \textbf{0.806044} & 0.818748 & \textbf{0.005874} & \textbf{0.4241} & \textbf{0.137099} \\
            Binning with PS & 0.804379  & 0.817290 & 0.007045  & 0.425184 & 0.137762 \\
        \bottomrule \\
        \end{tabular}
\end{center}
\caption{Calibration methods comparison on the loan-approval dataset.}
\label{tbl:calibrationEval2}
\end{table}

\subsection{Rain-forecasting domain}
\label{subsec:rainDetails}
For the rain-forecasting domain, we use a rain forecasting dataset \footnote{\url{https://www.kaggle.com/datasets/jsphyg/weather-dataset-rattle-package}}. This dataset was selected since it offers easily comprehensible predictions, making it accessible even to non-experts. This dataset spans approximately a decade of daily weather observations from numerous Australian weather stations. Australia, being a country where rainfall occurs for nearly half of the year, presents an ideal setting for constructing a model based on such balanced data. In total, this dataset contains $145,461$ rows corresponding to days in specific regions, with $22$ distinct features detailing rain-related information. These features include elements such as date, geographical region, minimum and maximum temperatures, precipitation levels, wind speed and gust, hours of sunshine, etc. The target label for prediction is whether it will rain on the following day. The dataset source is from the government of Australia, Bureau of Meteorology \footnote{\url{http://www.bom.gov.au/climate/dwo/}}, and is licensed under CC BY 4.0. We divided the data randomly into $80\%$ training, $10\%$ validation, and $10\%$ test. Note that in this domain, the rain-related information was not presented to the participants, but only the system's prediction.

The architecture of the MLP model is composed of four hidden layers with $64$, $32$, $16$, and $8$ neurons, respectively. We used the rectified linear unit (ReLU) as the activation function \citep{rectified2013Zeiler} and a sigmoid function in the activation function of the output. 
The optimizer used is \textit{Adam}, which is a popular optimization algorithm for training neural networks \citep{Adam2015Diederik}. The initial learning rate was set to $0.0001$. A batch size of $32$ was set for each iteration of training. The training process spanned $2,000$ iterations, ensuring sufficient time for convergence to be achieved.

During the survey, participants imagined planning an outdoor activity in a season and region where historical rainfall frequency is roughly 50 \%. After seeing the system’s rain probability for a particular day, they indicated how likely they were to cancel the activity. 

Each of the five methods was tested by 30 distinct workers, yielding a total of 150 responses (86 men and 64 women; mean age 41). Every worker received \$1 for the task, which took on average 19.5 minutes.

\subsection{Loan-approval domain}
\label{subsec:loanDetails}
For the loan-approval domain, we use a loan default prediction dataset \footnote{\url{https://www.kaggle.com/datasets/nikhil1e9/loan-default}}. This dataset contains $255,347$ loan records described by $18$ borrower–loan features such as loan amount, term length, interest rate, borrower age, income, credit score, education, etc. The binary target label indicates whether a loan ultimately defaults. This dataset is licensed under CC0: Public Domain. The data were divided randomly into training (80\%), validation (10\%), and test (10\%) sets.

The MLP, which served as our predictive model, consisted of four fully connected hidden layers with 128, 64, and 32 neurons, respectively. Between each layer, we applied batch normalization, a 10\% dropout rate, and ReLU activations. We added dropout and batch normalization since the dataset was biased for paid loans. The output layer employed a sigmoid activation to produce default‐probability estimates. We trained the network on CPU using the Adam optimizer with an initial learning rate of $0.0001$, a batch size of $32$, and $2,000$ iterations to ensure convergence.

The participants assumed the role of bank loan officers. For each scenario, after viewing the loan details and borrower information, they rated, on a five-point scale, how likely they were to approve the loan, then they observed the system’s prediction of repayment likelihood and could revise their decision.

To incentivize participants to pay attention to the data and the model’s predictions and to make sure their ratings were meaningful, we gave them a bonus or penalty according to their performance, in addition to the base payment for the survey. Here’s how the bonus worked: if the participant approved the loan (rated it $4$ or $5$) and the borrower repaid it, they received a bonus of $\$0.10$ if they rated it $5$ and $\$0.05$ if they rated it $4$. If the participant approved the loan and the borrower did not repay it, they incurred a penalty: $\$-0.10$ if they rated it $5$ and $\$-0.05$ if they rated it $4$. If the participant denied the loan (rated it $1$ or $2$) and the borrower did not repay it, they will receive a bonus: $\$0.10$ if they rated it $1$ and $\$0.05$ if they rated it $2$. If the participant denied the loan and the borrower repaid it, they incurred a penalty: $\$-0.10$ if they rated it $1$ and $\$-0.05$ if they rated it $2$.
If the participant rated the loan $3$, they will not receive a bonus or penalty.

Each of the five methods was presented to $35$ different participants, resulting in a total survey sample of $175$ participants. Among the $175$ participants, there are $99$ males, $75$ females, and $1$ person who chose not to identify, with an average age of $35.4$ years. 
Each participant received a base reward of $\$2$ for completing all the $20$ scenarios. The mean bonus was $\$0.31$.

\section{Results}

We evaluate our results using two primary criteria: (1) the average trust ratings assigned to each method, and (2) the average correlation between the model's predictions and participants' action ratings.  
All statistical significance tests reported in this section were conducted using pairwise one-way ANOVA comparisons across all methods.

\begin{figure}[htb]
    \centering
    \begin{subfigure}{0.49\linewidth}
        \centering
        \includegraphics[width=\linewidth]{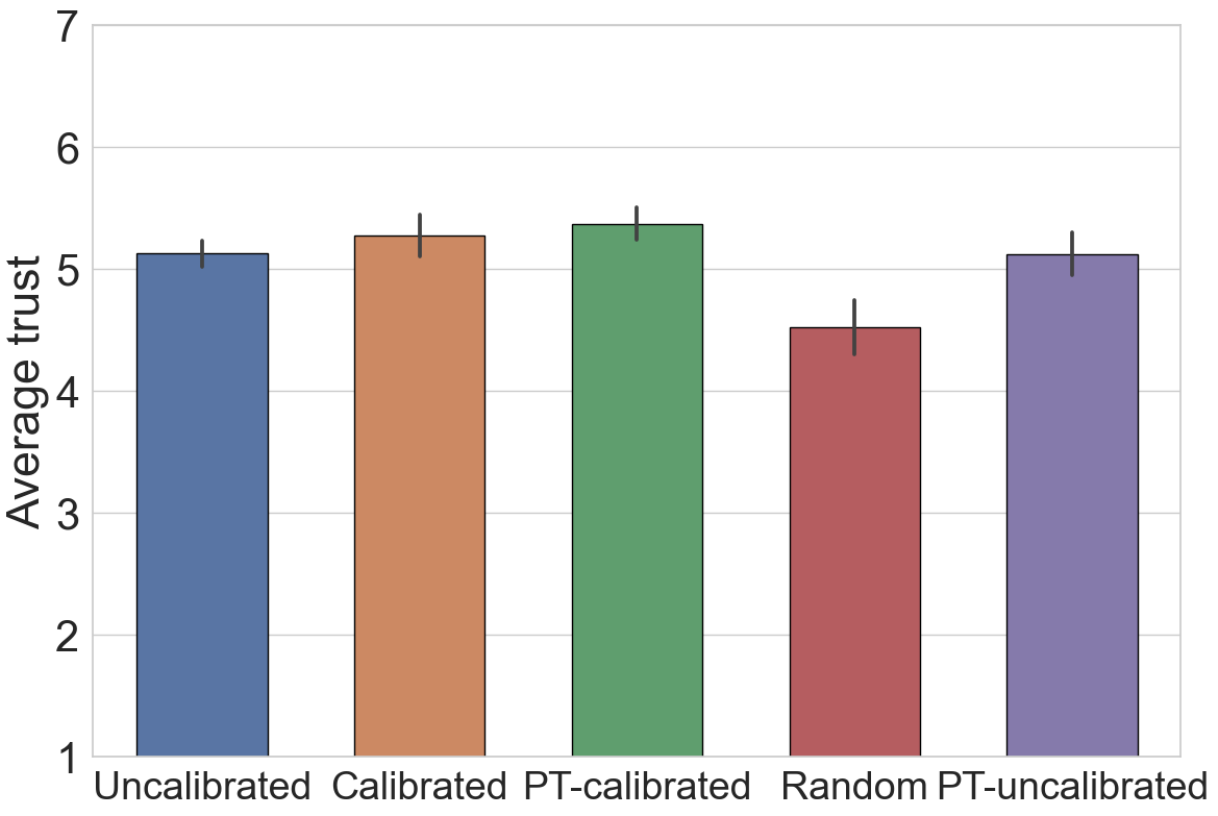}
        \caption{Average of all trust ratings in all $20$ scenarios. Error bars present the standard error.}
        \label{fig:resTrust1}
    \end{subfigure}
    \hfill
    \begin{subfigure}{0.49\linewidth}
        \centering
        \includegraphics[width=\linewidth]{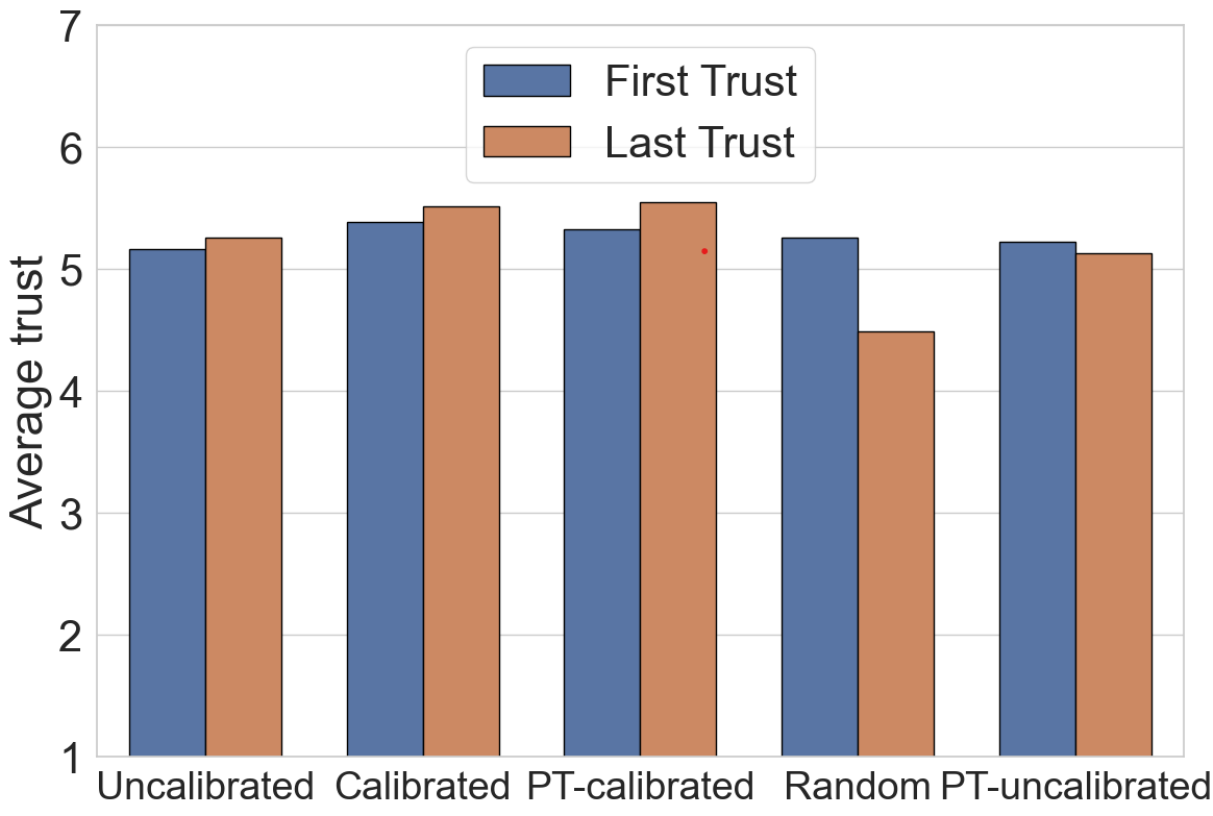}
        \caption{Average of trust ratings in the first and last scenarios only.}
        \label{fig:resTrust2}
    \end{subfigure}
    \caption{Average trust rating for each method in the rain-forecasting domain.}
    \label{fig:resTrust}
\end{figure}

\begin{figure}[htb]
    \centering
    \begin{subfigure}{0.49\linewidth}
        \centering
        \includegraphics[width=\linewidth, ]{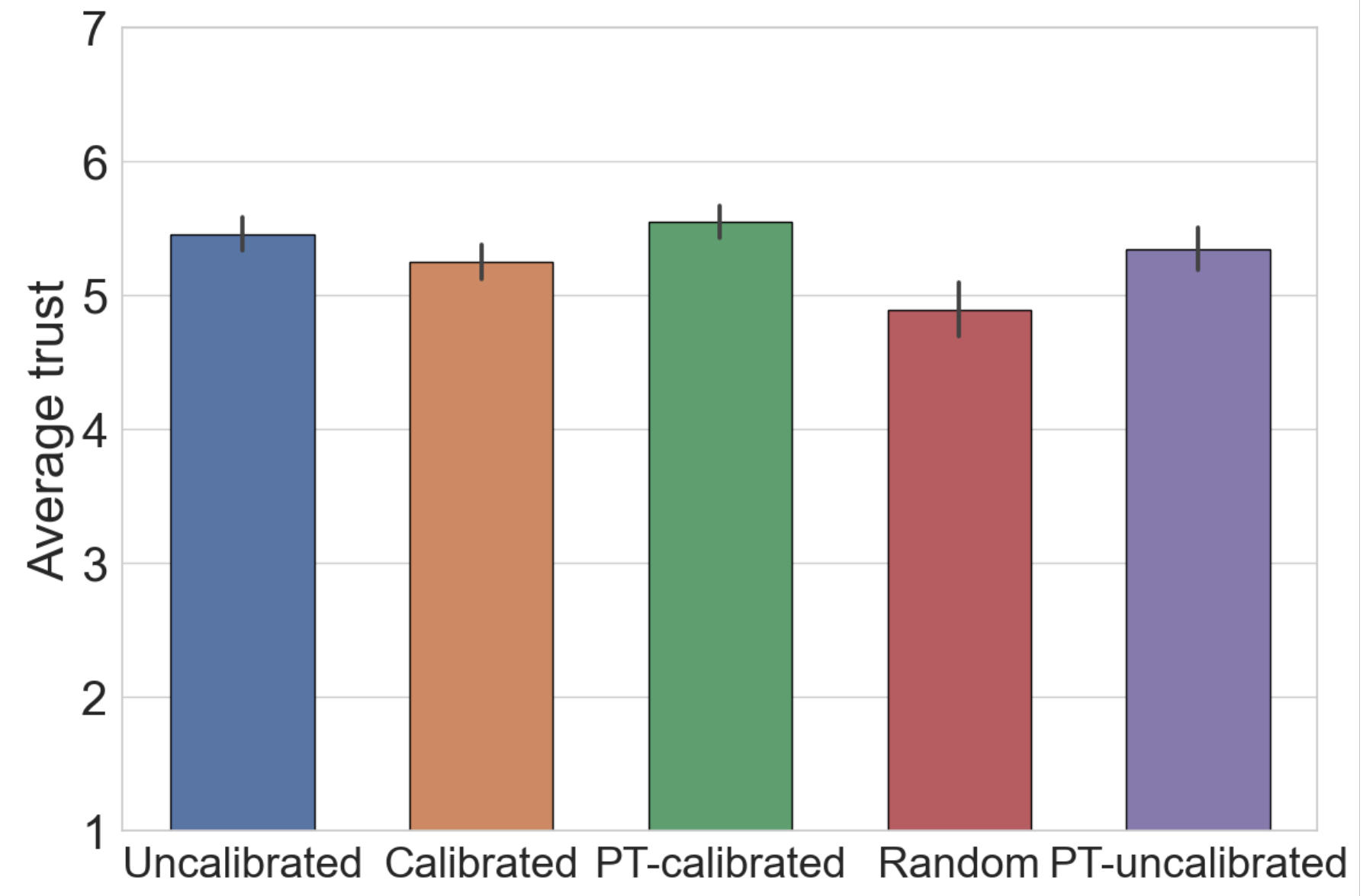}
        \caption{Average of all trust ratings in all $20$ scenarios. Error bars present the standard error.}
        \label{fig:resTrustLoan1}
    \end{subfigure}
    \hfill
    \begin{subfigure}{0.49\linewidth}
        \centering
        \includegraphics[width=\linewidth]{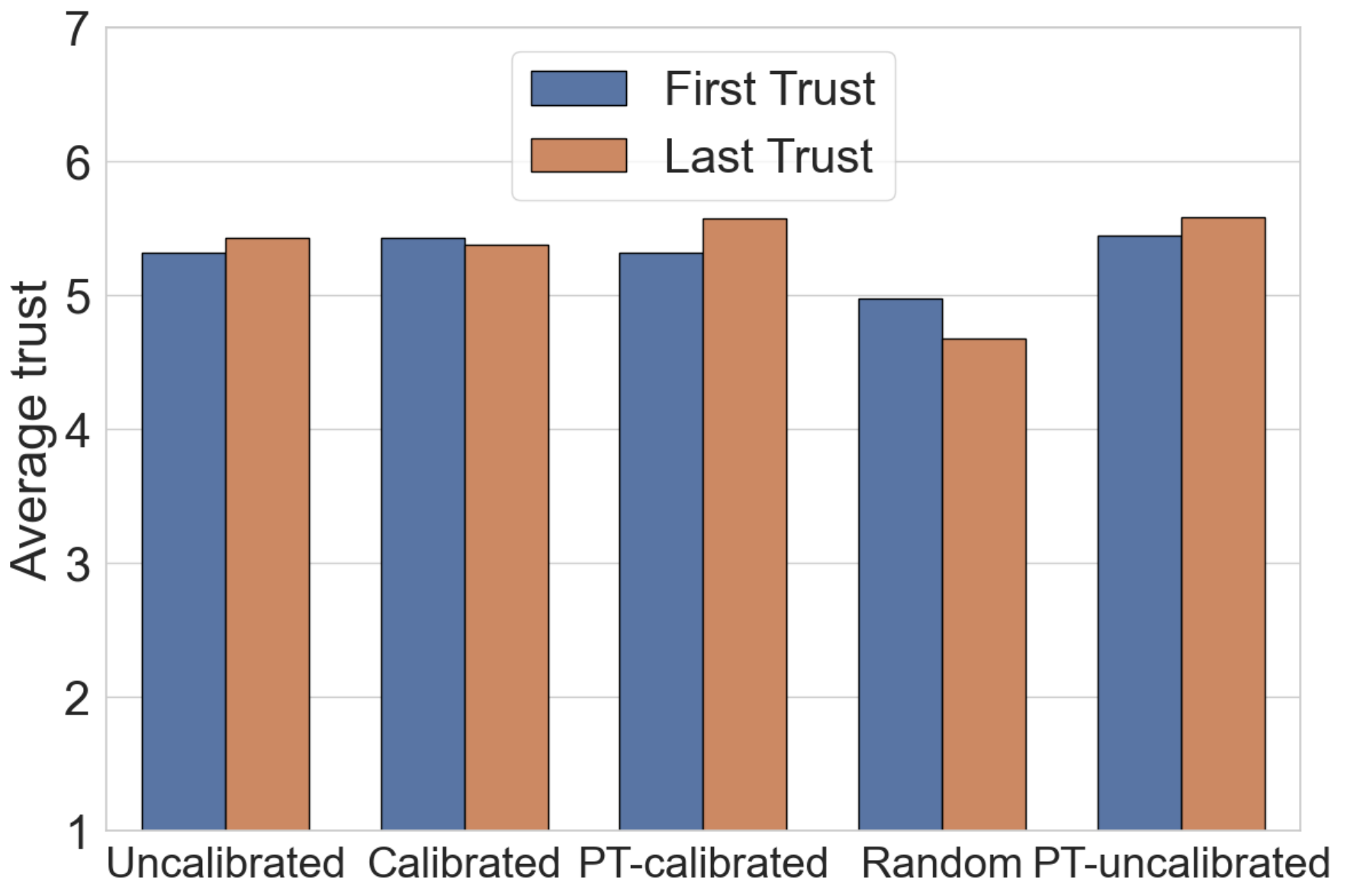}
        \caption{Average of trust ratings in the first and last scenarios only.}
        \label{fig:resTrustLoan2}
    \end{subfigure}
    \caption{Average trust rating for each method in the loan-approval domain.}
    \label{fig:resTrustLoan}
\end{figure}

Figures \ref{fig:resTrust1} and \ref{fig:resTrustLoan1} illustrate the average trust levels of participants across all scenarios for each method in the rain-forecast and the loan-approval domains, respectively. As anticipated, the random method received the lowest trust rating. However, when comparing the other methods, there is no notable difference in trust. In fact, except for the random model, there is no statistically significant difference between the models. Figures \ref{fig:resTrust2} and \ref{fig:resTrustLoan2} compare the trust levels in the first scenario to the trust level in the last scenario for each method in the rain-forecast and the loan-approval domains, respectively. Although there seems to be some increase in trust levels for some models and a decrease for others, these differences are not statistically significant.
Overall, it cannot be concluded that either calibration or the addition of a prospect theory layer increases people's trust in the model.

We now examine the correlation between the model's predictions and the participants' domain-specific actions. %likelihood of canceling the outdoor activity. 
%To that end, we compute the correlation between the model's predictions and the participant's likelihood of canceling the outdoor activity 
We compute the correlation for each participant and then average all the correlations for each method, in both domains. %MEIR LOOK HERE

\begin{figure}[htb]
    \begin{center}
    \includegraphics[width=0.7\linewidth]{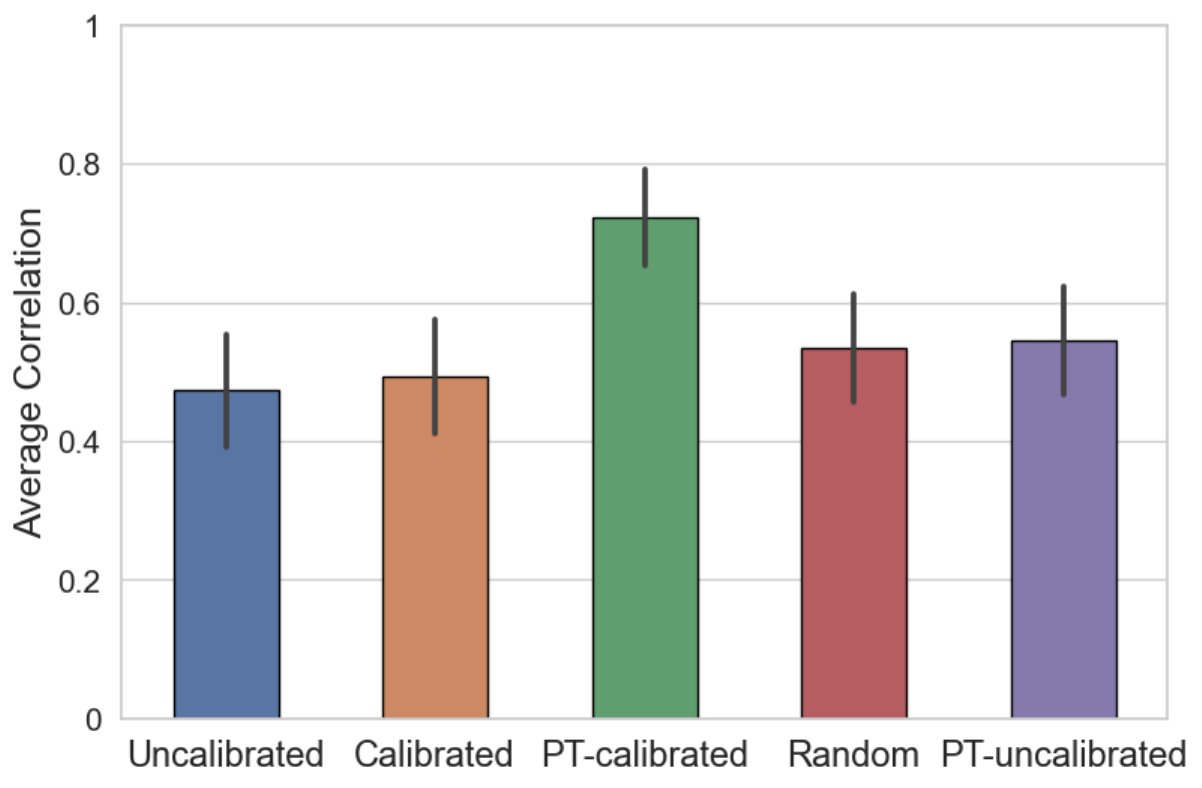}
    \end{center}
    \caption{Average correlation between the method predictions and the participants' rating to cancel the outdoor activity with Pearson correlation confidence interval (CI). The PT-calibrated method leads to a statistically significant increase in correlation compared to all other methods, in particular, the calibrated method.}
    \label{fig:resCorrelation}
\end{figure}

\begin{figure}[htb]
    \begin{center}
    \includegraphics[width=0.7\linewidth]{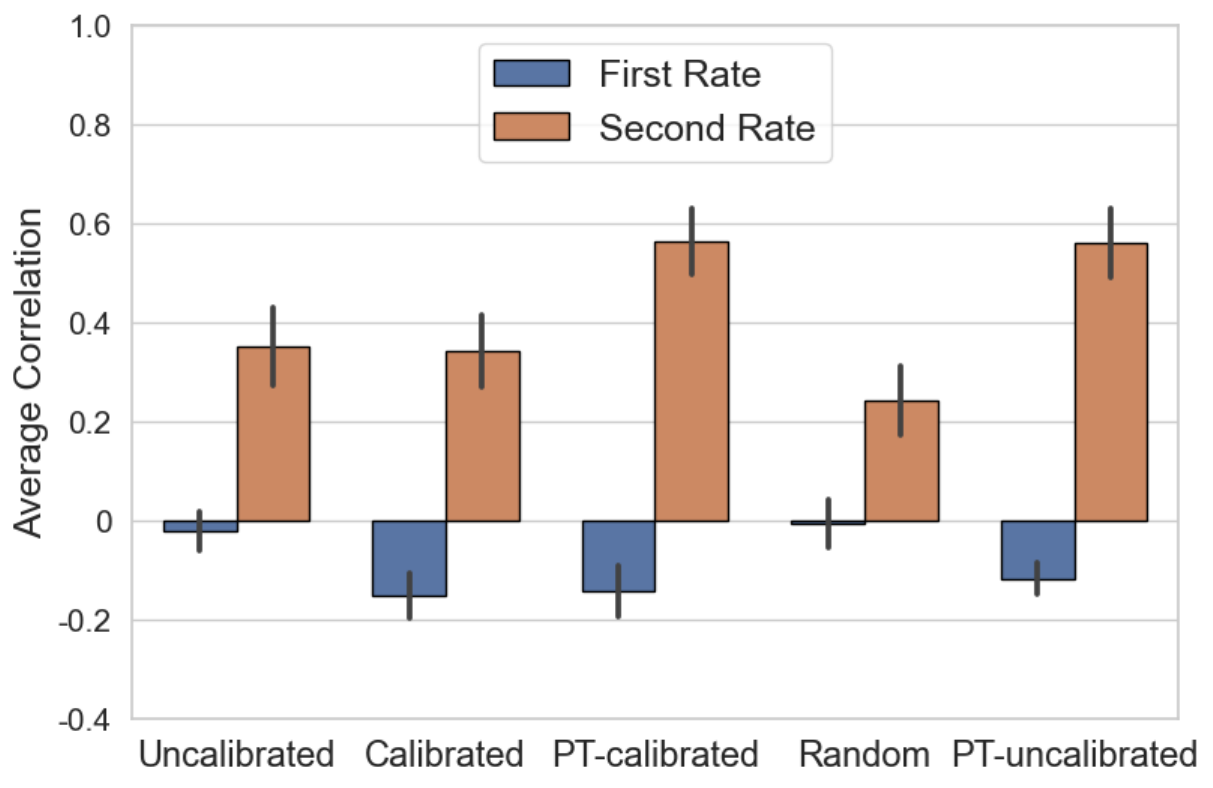}
    \end{center}
    \caption{Average correlation between the system's predictions and the participant's first rating to approve the loan and the participant's second rating after also seeing the system's prediction. The PT-calibrated method leads to a statistically significant increase in correlation compared to all other models, except the PT-uncalibrated model.}
    \label{fig:resCorrelationLoan}
\end{figure}

As depicted by Figures \ref{fig:resCorrelation} and \ref{fig:resCorrelationLoan}, there are notable differences between the models. Specifically, the correlation observed for the PT-calibrated model is higher than all other methods in both domains. These differences are statistically significant for all pairwise comparisons ($p<0.05$) except when compared to the PT-uncalibrated method in the loan-approval domain. Interestingly, incorporating the prospect theory probability adjustments into the uncalibrated model significantly improved the correlation only in the rain-forecast domain, while in the loan-approval domain, the improvement was negligible. %It can be inferred that applying prospect theory independently, without calibrating the model, yields some improvement, though not statistically significant, and it even lowers participants' trust in the model.

Furthermore, when the final outcomes are random and unrelated to the model's predictions, the correlation is significantly lower (i.e., the random method shows a significantly lower correlation than the PT-calibrated model in both domains). %However, it still yields a higher correlation than both the calibrated and uncalibrated models. We believe 
This indicates that the observed differences between the PT-calibrated method and the other methods are not solely due to the reported probabilities, but are in relation to the true predictions.

In addition, as illustrated in the figures, the calibrated method performed very similarly to the uncalibrated method, showing a marginal improvement in the rain-forecasting domain and a slight degradation in the loan-approval domain.
This implies that calibrating the model alone does not necessarily influence people to base their decisions on its predictions. However, adding to the calibrated model a layer that adjusts probabilities in line with people's expectations, as per prospect theory, encourages individuals to align their decisions with the model's predictions. 

% -----------------------------------------------------------

\section{Conclusion and plans for further research}
\label{sec:conclusion}
In this paper, we study how people react to the probabilities predicted by a machine learning model. We explore whether calibrating the probabilities increases trust in the system and whether it leads to people making decisions that are more aligned with the system's predictions. Furthermore, we introduce a prospect theory-based correction that is applied to the model's calibrated probabilities. We show that while the trust in the system is not significantly affected by the method used, when asked to take action, the resulting correlation between the model's prediction and the human action is significantly higher for the model with calibration and prospect theory correction. Notably, we also observe that calibration alone does not necessarily improve alignment between human decisions and the system's prediction. These results were obtained in two different domains.

 %Consequently, we recognize the need to explore alternative and innovative approaches focused on bolstering user trust. One promising avenue involves leveraging reinforcement learning techniques to train models explicitly designed to enhance user trust in the predictions they provide. This exploration could open up exciting possibilities for improving the acceptance and reliability of machine learning models in practical applications.

One limitation of our work is that we used only two domains. In future work, we will consider testing additional domains such as recidivism, which refers to the relapse into criminal behavior. That is, we would present criminal information to a participant and expect her to predict if the criminal will relapse into criminal behavior in the future, before and after observing the system's prediction. 
Another limitation of our work is that the participants did not perform any action in the real world. While we tried to mitigate this slightly in the loan-approval domain, in which they obtained a bonus depending on their performance, the behavior might differ when the stacks are much higher in a real-world scenario.

% Future Work ideas
A third limitation of our study is that we use a standard $\gamma$ value for the prospect theory correction. This $\gamma$ value is based on a domain-agnostic international survey. Estimating $\gamma$ precisely for the domain or population of interest (rain forecasting in Australia, for instance) is of natural interest. In future work, we intend to study whether using domain-specific $\gamma$ further increases the effect size in our findings. Such an approach would dynamically adjust the $\gamma$ values within the inverse weighting function to align with the user preferences. %The objective of implementing such a system is to empower users to make decisions that align closely with the predictions generated by the machine learning model.

% -----------------------------------------------------------
\clearpage

\bibliographystyle{chicago}
\bibliography{main}

% -----------------------------------------------------------

\clearpage
\appendix
\section*{Appendix}
\label{subsec:cal}

\section{Experiments screenshots}
\begin{figure}[htb]
\centering
    \begin{center}
        \includegraphics[width=\linewidth, height=13cm]{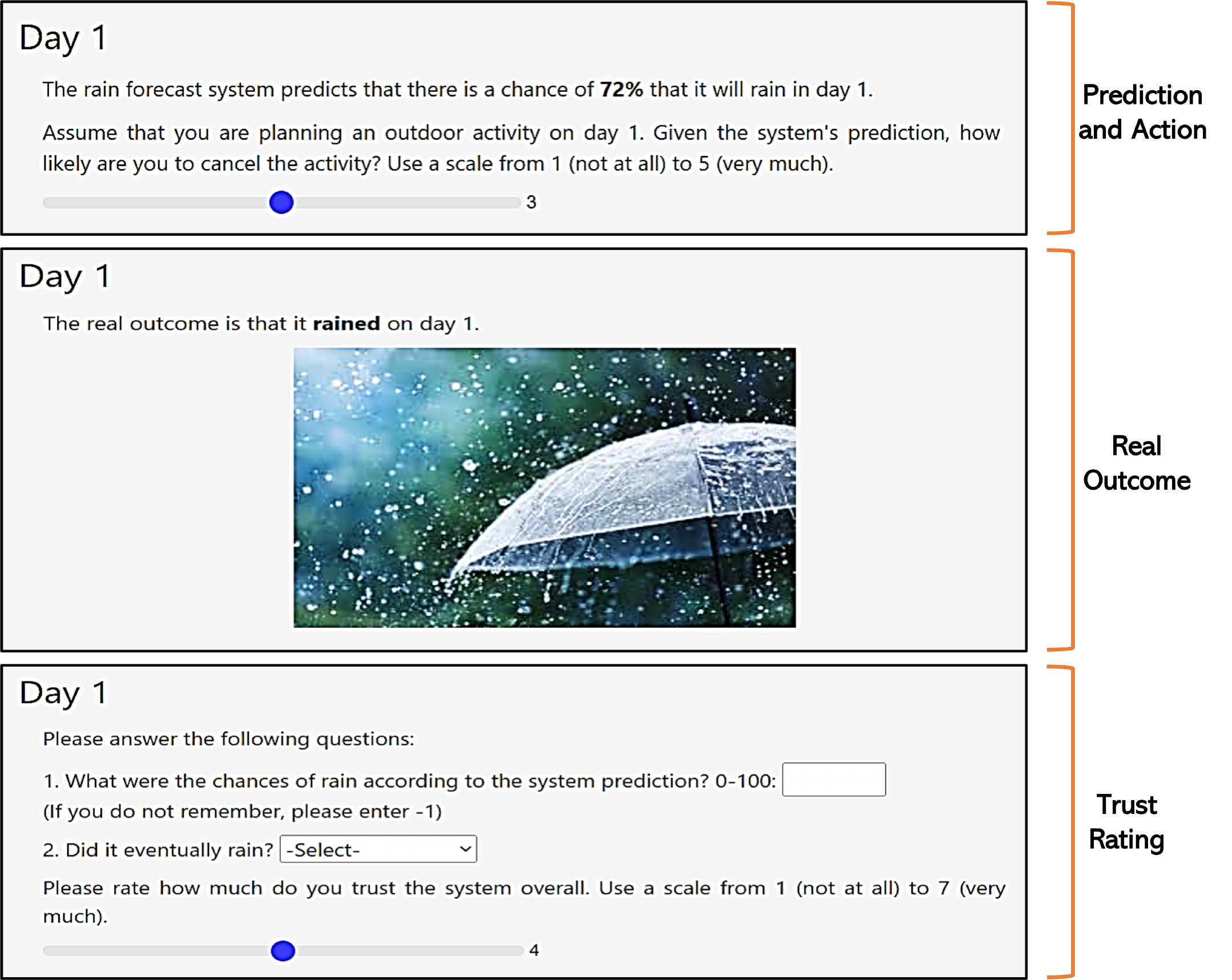}
    \end{center}
    \caption{Interface of the experimental task in the survey.}
    \label{fig:survey}
\end{figure}

Figure \ref{fig:survey} is an example of these rain-forecasting experiment for the first day.

\begin{figure}[htb]
    \begin{center}
        \includegraphics[width=\linewidth]{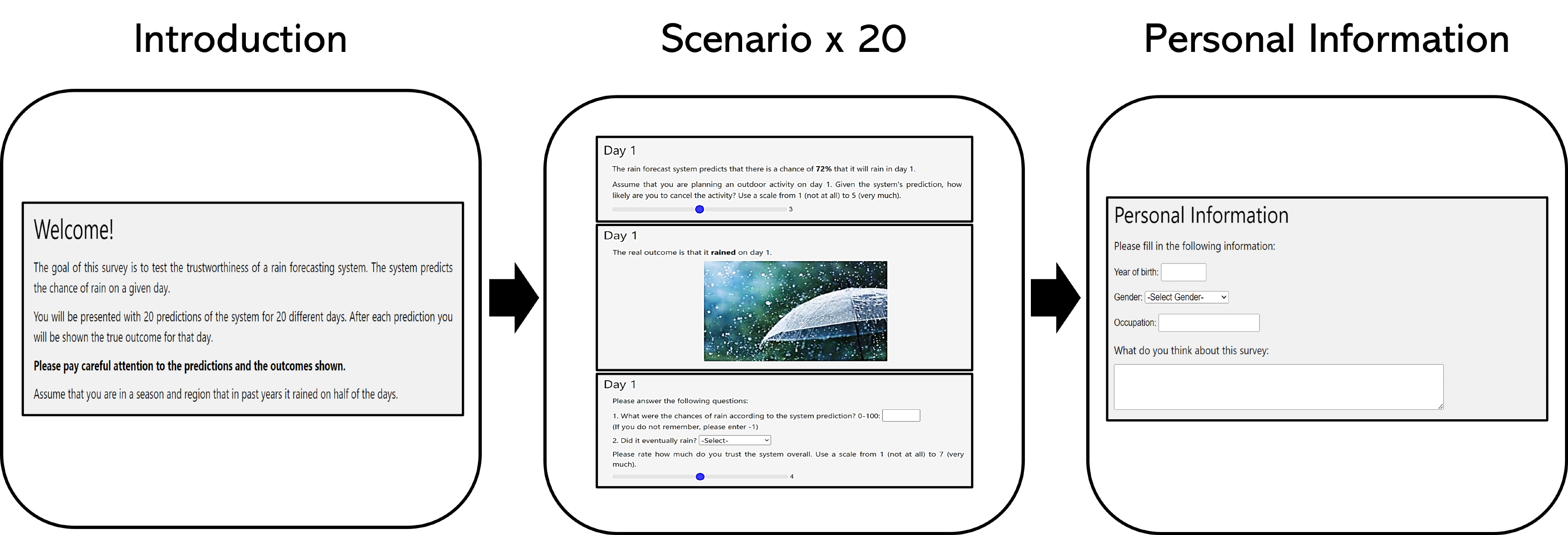}
    \end{center}
    \caption{Flowchart of the survey.} 
    \label{fig:surveyFlow}
\end{figure}

Figure \ref{fig:surveyFlow} depicts the flow of the entire survey for each participant.

% model calibrations methods
\section{Calibration methods}
\label{subsec:cal-met}
The following are common methods to calibrate a model. All methods are post-processing steps that produce calibrated probabilities. That is, given a model's output $p_i$ on sample $x_i$, they produce a calibrated probability $q_i$. To avoid unwanted bias, each method requires a validation set, which can also be used for hyperparameter tuning.

\begin{enumerate}

\item {\bf Platt scaling} \citep{platt1999probabilistic}. A parametric approach that assumes a sigmoidal relationship between the model's outputs and the true probabilities. It trains a logistic regression model on the validation set, where the original model's outputs are used as features. That is, it learns scalars $a, b \in \mathbb{R}$ and outputs $q_i=\sigma(ap_i+b)$ as the calibrated probability.
Platt scaling is particularly effective for max-margin methods such as SVMs and boosted trees, which show sigmoidal distortions in their predicted probabilities \citep{niculescu2005predicting}.

Platt scaling can be extended to multiclass models by applying to the model's logits vector, $z_i$,  a linear transformation $Az_i + b$, where $A \in \mathbb{R}^{K \times K}$ and $b \in \mathbb{R}$. The calibrated probability of the predicted class is $q_i=\max_k \sigma \left(Az_i + b \right)^{(k)}$ \citep{guo2017calibration}. 

\item {\bf Binning} \citep{zadrozny2001obtaining}. A simple non-parametric approach. It sorts all the model's predictions and divides them into $M$ interval bins. The bin boundaries are either chosen to be equal length intervals or to equalize the number of samples in each bin. Given a prediction $p_i$,
the method finds the bin containing that prediction and returns as output the fraction of positive outcomes in the bin. This method has several limitations, including the need to define the number of bins and the fact that the bins and their associated boundaries remain fixed on all predictions.
\citep{naeini2015obtaining} proposed BBQ, a refinement of the binning method using Bayesian model averaging.

\item {\bf Isotonic regression} \citep{zadrozny2002transforming}. A more general non-parametric approach that only assumes a monotonic increasing relationship between the model's outputs and the true probabilities. It utilizes isotonic regression \citep{robertson1988order} to learn an isotonic (non-decreasing) function $f$ to map $q_i=f(p_i)$. A common algorithm used for computing the isotonic regression model is the PAV algorithm \citep{ayer1955empirical}.
Isotonic regression is a more powerful calibration method that can correct any monotonic distortion; however, it is more prone to overfitting \citep{niculescu2005predicting}.

Isotonic regression can be extended to multiclass models by reducing the multiclass problem into a set of binary problems. A well-known approach to this end is one-against-all, in which a classifier is trained for each class using as positives the examples that belong to that class, and as negatives all other examples. At test time, we obtain probability vector $[q_i^{(1)},...,q_i^{(K)}]$ where $q_i^{(k)}$ is the calibrated probability for class $k$. The final prediction probability $q_i$ is the max of the vector normalized by $\sum_{k=1}^{K}{q_i^{(k)}}$.
Another common approach is all-pairs \citep{allwein2000reducing}.

\item {\bf Temperature scaling} \citep{guo2017calibration}. A simple extension of Platt scaling that learns a single Temperature parameter $t > 0$ to rescale the logit vector $z_i$. The calibrated probability of the predicted class is $q_i=\max_k \sigma \left(z_i/t \right)^{(k)}$. Temperature scaling does not affect the model’s accuracy as $t$ does not change the maximum of the softmax function.

\item {\bf Ante-hoc methods}.
An alternative to post-hoc calibration is to modify the classifier learning algorithm itself. MMCE \citep{kumar2018trainable} trains neural networks by optimizing the combination of log loss with a kernel-based measure of calibration loss. SWAG \citep{Maddox2019Simple} models the posterior distribution over the weights of the neural network and then samples from this distribution to perform Bayesian model averaging. \citep{milios2018dirichlet} proposed a method to transform the classification task into regression and to learn a Gaussian Process model.

\end{enumerate}

% ---------------------------------------------------------

\section{Evaluating calibration}
\label{subsec:cal-eval}
% Notations to evaluate model calibration
To evaluate the calibration of the model from a finite set of $n$ samples, the predicted probabilities are grouped into $M$ interval bins of equal size. Let $B_m$ be the set of samples whose predicted probability falls into bin $m$. The accuracy of $B_m$ is $$acc(B_m)=\frac{1}{|B_m|} \sum_{i \in B_m}{1(\hat{y_i}=y_i)},$$ where $\hat{y_i}$ and $y_i$ are the predicted and true class labels for sample $i$. The average confidence of $B_m$ is $$conf(B_m)=\frac{1}{|B_m|} \sum_{i \in B_m}{\hat{p_i}},$$ where $\hat{p_i}$ is the confidence score the model assigns to sample $i$ for belonging to class $y_i$. A perfectly calibrated model will have $acc(B_m) = conf(B_m)$ for all $m \in [M]$.

% methods to evaluate calibration
The following are common methods to evaluate the calibration of a model.
\begin{enumerate}

\item {\bf Reliability Diagram} \citep{murphy1977reliability}: is a visual representation of model calibration. These diagrams plot for every bin $m \in [M]$ the accuracy of $B_m$ as a function of the confidence of $B_m$. If the model is perfectly calibrated, then the diagram should plot the identity function. Any deviation from a perfect diagonal represents miscalibration. When the diagram is above the diagonal, the model is under-predicting the true probability, and if it is below, the model is over-predicting the true probability. %See figure \ref{fig:reliabilityDiagram} for an example of a reliability diagram.

\item {\bf Expected Calibration Error (ECE)} \citep{naeini2015obtaining}. While reliability diagrams are useful visual tools, it is more convenient to have a scalar summary statistic of calibration. ECE approximates the difference in expectation between confidence and accuracy by taking a weighted average of the bins’ accuracy/confidence difference. $$ECE = \sum_{m=1}^{M}{\frac{|B_m|}{n} \bigg|acc(B_m)-conf(B_m)\bigg|}$$ 
A well-calibrated model will have a small ECE, indicating that its predicted probabilities closely reflect its actual performance, whereas a poorly calibrated model will have a large ECE.

\item {\bf Maximum Calibration Error (MCE) }\citep{naeini2015obtaining}. In high-risk applications, confident but wrong predictions can be especially harmful. In such cases, we may wish to minimize the worst-case deviation between confidence and accuracy. MCE estimates an upper bound of this deviation. 
$$MCE = max_{1 \in [M]}{\big|acc(B_m)-conf(B_m)\big|}$$ 

\item {\bf Overconfidence Error (OE)}. Another method that measures the overconfidence of a model. This metric penalizes predictions by the weight of the confidence, but only when confidence exceeds accuracy, and thus overconfident bins incur a high penalty.
$$OE = \sum_{m=1}^{M}{\frac{|B_m|}{n} \bigl[conf(B_m) \cdot max \left(conf(B_m)-acc(B_m),0\right)\bigr]}$$

\item {\bf Negative log-likelihood (NLL)}. Measures the difference between the confidence score and the true probabilities of the outcomes, as expressed in the likelihood function.
$$NLL = -\sum_{i=1}^{n}{\log(\hat{p_i}})$$
The lower the NLL, the better the calibration of the model. Note that the negative log-likelihood is also equivalent to the cross-entropy loss, which is commonly used in machine learning for classification tasks \citep{de2005tutorial}. 

\item {\bf Brier Score (BS)} \citep{brier1950verification}. is the mean squared error between the confidence scores and the true class labels.
$$BS = \frac{1}{n} \sum_{i=1}^{n}{ \sum_{k=1}^{K}(y_i^{(k)}-p_i^{(k)})^2},$$
where $K$ is the number of classes, $y_i^{(k)}$ is a binary variable that indicates whether the sample $x_i$ belongs to class $k$ and $p_i^{(k)}$ is the probability the model assigns to sample $x_i$ for belonging to class $k$. The Brier score ranges from 0 to 1, where 0 indicates perfect calibration and 1 indicates the worst possible calibration.

\end{enumerate}

\end{document}